\documentclass[aps,prb,twocolumn,superscriptaddress]{revtex4-1}
\usepackage{graphicx}
\usepackage{float}

\usepackage{amsmath}
\newcommand{\be}{\begin{equation}}
\newcommand{\ee}{\end{equation}}
\newcommand{\bi}{\begin{itemize}}
\newcommand{\ei}{\end{itemize}}
\newcommand{\bea}{\begin{eqnarray}}
\newcommand{\eea}{\end{eqnarray}}
\newcommand{\Eq}[1]{Eq.(\ref{#1})}
\newcommand{\Fig}[1]{Fig.\,\ref{#1}}
\newcommand{\Tab}[1]{Table \,\ref{#1}}
\newcommand{\Onlinecite}[1]{Ref.\,\onlinecite{#1}} %
\newcommand{\Xhh}{\ensuremath{\rm X_{hh}}}

\newcommand{\Xlh}{\ensuremath{\rm X_{lh}}}
\newcommand{\Xso}{\ensuremath{\rm X_{so}}}
\newcommand{\Chh}{\ensuremath{\rm C_{hh}}}
\newcommand{\Clh}{\ensuremath{\rm C_{lh}}}
\newcommand{\Rhh}{\ensuremath{R_{\rm hh}}}
\newcommand{\Rlh}{\ensuremath{R_{\rm lh}}}
\newcommand{\mhh}{\ensuremath{m_{\rm hh}}}
\newcommand{\mlh}{\ensuremath{m_{\rm lh}}}
\newcommand{\mc}{\ensuremath{m_{\rm c}}}
\newcommand{\muhh}{\ensuremath{\mu_{\rm hh}}}
\newcommand{\mulh}{\ensuremath{\mu_{\rm lh}}}
\newcommand{\me}{\ensuremath{m_{\rm e}}}
\newcommand{\deltaDB}{\ensuremath{\delta_{0}}}
\newcommand{\omp}{\ensuremath{\omega_{\rm p}}}
\newcommand{\gp}{\ensuremath{\gamma_{\rm p}}}
\DeclareMathOperator\erf{erf}

\newcommand{\aB}{\ensuremath{a_{\rm B}}}

\newcommand{\aext}{\ensuremath{\alpha_{\rm ext}}}
\newcommand{\pX}{\ensuremath{p_{\rm X}}}
\newcommand{\AC}{\ensuremath{A_{\rm C}}}
\newcommand{\omB}{\ensuremath{\omega_{\rm B}}}
\newcommand{\gC}{\ensuremath{\gamma_{\rm C}}}
\newcommand{\gL}{\ensuremath{\gamma_{\rm L}}}

\begin{document}


\title{Evidence of giant oscillator strength in the exciton dephasing of CdSe nanoplatelets measured by resonant four-wave mixing}


\author{Ali Naeem}
\affiliation{Cardiff University School of Physics and Astronomy, The Parade, Cardiff CF24 3AA, United Kingdom}
\author{Francesco Masia}
\affiliation{Cardiff University School of Physics and Astronomy, The Parade, Cardiff CF24 3AA, United Kingdom}
\author{Sotirios Christodoulou}
\affiliation{Istituto Italiano di Tecnologia, Via Morego 30, IT-16163 Genova, Italy}
\author{Iwan Moreels}
\affiliation{Istituto Italiano di Tecnologia, Via Morego 30, IT-16163 Genova, Italy}
\author{Paola Borri}
\affiliation{Cardiff University School
of Physics and Astronomy, The Parade, Cardiff CF24 3AA, United Kingdom}
\affiliation{Cardiff University School of Biosciences, Museum Avenue, Cardiff CF10 3AX, United Kingdom}

\author{Wolfgang Langbein}
\email{langbeinww@cardiff.ac.uk}
\affiliation{Cardiff University School of Physics and Astronomy, The Parade, Cardiff CF24 3AA, United Kingdom}


\date{\today}

\begin{abstract}
We measured the intrinsic ground-state exciton dephasing and population dynamics in colloidal quasi two-dimensional (2D) CdSe nanoplatelets at low temperature (5-50\,K) using transient resonant four-wave mixing in heterodyne detection. Our results indicate that below 20\,K the exciton dephasing is lifetime limited, with the exciton population lifetime being as fast as 1\,ps. This is consistent with an exciton lifetime given by a fast radiative decay due to the large in-plane coherence area of the exciton center-of-mass motion in these quasi 2D systems compared to spherical nanocrystals.
\end{abstract}



\maketitle

The colloidal synthesis of quasi-2D semiconductor nanostructures has recently attracted much attention, owing to the simplicity, flexibility and low cost of colloidal chemistry compared to epitaxial growth techniques, and the wealth of interesting fundamental properties and applications of quantum wells (QWs) in e.g. optoelectronics and photovoltaics. High--quality colloidal zinc-blende CdSe nanoplatelets (NPLs) having a thickness of 1-2\,nm were recently reported\,\cite{IthurriaJACS08,IthurriaNMa11,TessierACSN12}, and exhibit absorbtion spectra well described by a QW-like electronic structure. Remarkably, the synthesized ensembles can have a monolayer thickness purity better than 95\%, and the inhomogeneous broadening corresponds to only about 20\% of the monolayer splitting, which is similar to optimized epitaxial quantum wells \cite{LeossonPRB00}. Furthermore, the thickness quantization energy of 0.5-1\,eV is much larger than the bulk exciton binding energy of 15\,meV \cite{VoigtPSSB79}, such that the excitons are close to the 2D limit providing a fourfold binding energy increase. The binding energy is further enhanced \cite{MuljarovPRB95} by the lower dielectric constant $\varepsilon\sim2$ in the NPL surrounding, and the lower dielectric constant $\varepsilon_\infty \sim 6$ of CdSe for energies above the LO phonon energy of 26\,meV compared to $\varepsilon_{\rm s}\sim 10$ below, resulting in predicted exciton binding energies \cite{AchtsteinNL12} in the 100-300\,meV range.

Since the exciton oscillator strength increases with the exciton binding energy, we can expect a fast exciton radiative decay. Recent reports showed photoluminescence (PL) lifetimes decreasing with decreasing temperature, and lifetimes of $200-400$\,ps were measured at low temperatures\,\cite{IthurriaNMa11,TessierACSN12, AchtsteinNL12}, two orders of magnitude faster than in spherical CdSe nanocrystals.  It is understood theoretically and reported experimentally in epitaxially--grown QWs that QW excitons exhibit an oscillator strength which increases with increasing extension of the exciton in-plane center-of-mass (CM) motion wavefunction \,\cite{FeldmannPRL87,AndreaniPRB99,HoursPRB05,SavonaPRB06}, the so-called coherence area. We therefore expect that the fundamental bright NPL exciton (BX) has a short radiative lifetime decreasing with increasing NPL area.

The radiative decay rate also sets a lower limit to the homogeneous linewidth of an optical transition. Recent PL measurements in single NPLs at low temperature showed linewidths of 0.5-1\,meV\,\cite{TessierACSN12,AchtsteinNL12}. These would correspond to a population lifetime in the 1\,ps range, significantly shorter than the measured PL decay time. It is however known that single quantum dot PL linewidths are affected by fluctuations of the emission energy during long acquisition times (so-called spectral diffusion), hence the reported linewidths give an upper limit to the homogeneous linewidth. Furthermore, the PL decay of non-resonantly excited platelets does not provide a measurement of the BX lifetime but reflects the density dynamics mediated by phonon-scattering across all occupied exciton states in the NPL, including higher CM quantized exciton states of lower oscillator strength, and spin-forbidden dark states. The measured PL decay time is thus only an upper limit of the BX decay time. To investigate the giant oscillator strength effect in NPL it is therefore important to measure the intrinsic homogeneous linewidth and lifetime of the BX.

We previously demonstrated both in epitaxial quantum wells \cite{BorriPRB99,LangbeinPRB00}, self-assembled quantum dots \cite{BorriPRL01,BorriPRB05} and colloidal nanocrystals \cite{MasiaPRB11,MasiaPRL12,AccantoACSN12} that transient resonant four-wave mixing (FWM) can measure the intrinsic exciton dephasing time in inhomogeneously broadened ensembles unaffected by spectral diffusion. In the present work, we have used three-beam FWM to measure the intrinsic exciton dephasing and population dynamics in colloidal zinc-blende CdSe NPLs in the temperature range from 5\,K to 50\,K.

The investigated NPLs have been synthesized according to the method reported in \Onlinecite{IthurriaNMa11}, albeit using a twice larger Se concentration, and injecting 2.25 times more cadmium acetate, at a temperature of 210$^\circ$C. The resulting NPL have a room temperature emission around 515\,nm, and the X-ray diffraction reveals that they possess a zinc-blende crystal structure. More details are given in the supplement.

\begin{figure}[t]
\centerline{\includegraphics*[width=8cm]{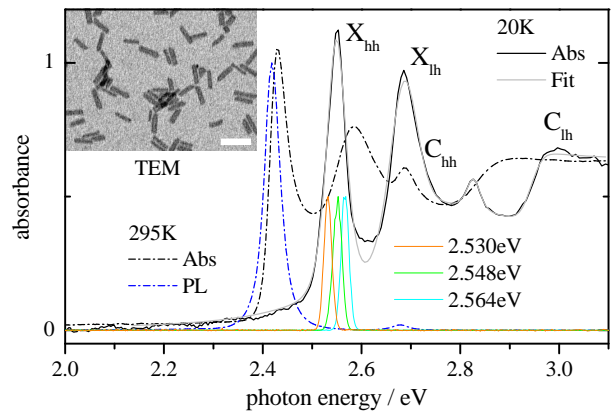}}
\caption{Linear optical properties of the investigated CdSe NPL ensemble. Absorption and photoluminescence spectra at 295\,K (dash-dotted lines), and absorbtion at 20\,K with a fit (solid lines). The spectra of the laser pulses used in the FWM experiment are also shown, labeled according to their center photon energy. Inset: TEM image of the NPLs. Scale bar: 50\,nm. \label{fig:AbsTEM}}
\end{figure}

An optical and structural characterization of the investigated NPL ensemble is shown in \Fig{fig:AbsTEM}. The synthesis typically yields NPLs with a room-temperature PL quantum efficiency of 50\% and PL lifetimes in the nanosecond range, indicating that the non-radiative decay due to defects is slow. Transmission electron microscopy shows NPLs with lateral dimensions of  $L_x=30.8\pm2.6$\,nm and $L_y=7.1\pm0.9$\,nm. Room-temperature absorption and emission spectra of the ensemble reveal a small Stokes shift (12\,meV), and a wavelength of the lowest excitonic transition \Xhh\ of 486.5\,nm consistent with an electron/heavy-hole exciton confined in 6 monolayer (ML) thickness ($L_z=1.82$\,nm) according to \Onlinecite{IthurriaNMa11}. The small absorption and emission peak observed at 2.69\,eV (460\,nm) reveals a few percent of NPLs with 5ML thickness in the ensemble, which are not affecting the FWM in resonance with \Xhh\ due to the large energy shift. After cooling to 20\,K the absorption spectrum shifts to higher energies and exhibits a narrower \Xhh\ due to the reduction of the phonon-scattering related homogenous broadening \cite{AchtsteinNL12,TessierACSN12}. The absorption lineshape at low temperature was fitted by a sum of two excitonic peaks \Xhh, \Xlh\ and continuum edges \Chh,\Clh, plus an additional peak for the 5ML contribution (for details see supplement). We inferred a \Xhh\ linewidth of $(46\pm 1)$\,meV full-width at half-maximum (FWHM) dominated by inhomogeneous broadening.
Furthermore, excitonic binding energies of $\Rhh=(178\pm34)\,$meV and $\Rlh=(259\pm3)\,$meV are inferred from the fit. \Rhh\ is consistent with the range of $100-300$\,meV predicted in calculations \cite{AchtsteinNL12}. The difference of \Rhh\ and \Rlh\ can be attributed to the different in-plane hole dispersions, as the 2D exciton binding energy is proportional to the in-plane reduced mass. Notably, the heavy-hole quantized by the NPL thickness has in-plane a light-hole mass $\mlh=0.19\,\me$, and the light-hole has in-plane a heavy-hole mass $\mhh=0.67\,\me$, as deduced from the Pidgeon-Brown model used in \Onlinecite{IthurriaNMa11}. With the isotropic electron mass of $\mc=0.18\,\me$, this results in an in-plane reduced mass of $\muhh=0.092\,\me$ for \Xhh\ and $\mulh=0.14\,\me$ for \Xlh. The expected ratio of binding energies is thus $\Rlh/\Rhh=\mulh/\muhh=1.51$, which is close to the ratio of 1.4 inferred from the fit.

\begin{figure}[t]
\centerline{\includegraphics*[width=8cm]{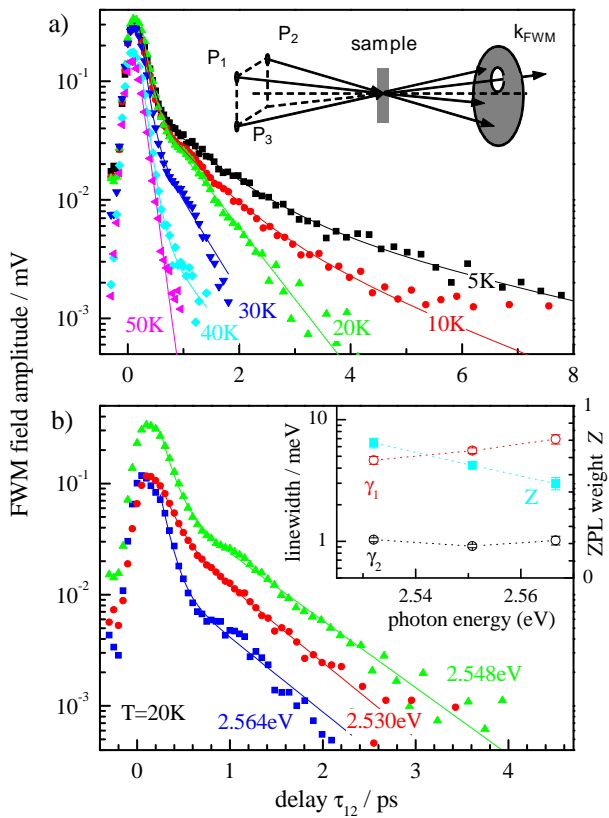}}
\caption{FWM field amplitude versus delay between the first two exciting pulses $\tau_{12}$ at $\tau_{23}=1$\,ps. a) At different temperatures as indicated for a center energy of 2.530\,eV. The lines are fits to the data. The inset shows a sketch of the directional selection geometry used in the experiment. b) For different center energies as indicated (spectra given in \Fig{fig:AbsTEM}) at a temperature of $20$\,K. The lines are fits of a bi-exponential decay to the data. The inset shows the resulting linewidths $2\hbar\gamma_{1,2}$. \label{fig:Dephasing}}
\end{figure}

Similar to previous works on CdSe nanocrystals\,\cite{MasiaPRL12,AccantoACSN12}, we have measured the dephasing time of the BX using transient three-beam FWM (see sketch in \Fig{fig:Dephasing}) in resonance with \Xhh\ (see laser spectra in \Fig{fig:AbsTEM}). All beams are derived from a train of 150\,fs pulses with 76\,MHz repetition rate. The first pulse ($P_1$) induces a coherent polarization in the sample, which after a delay $\tau_{12}$ is converted into a density grating by the second pulse ($P_2$). The third pulse ($P_3$), delayed by $\tau_{23}$ from $P_2$, is diffracted by this density grating, yielding the FWM signal. In the employed heterodyne technique \cite{BorriJPCM07} the pulse trains are radio-frequency shifted resulting in a frequency-shifted FWM field which is detected by its interference with a reference pulse. In an inhomogeneously broadened ensemble, the FWM signal is a photon echo emitted at a time $\tau_{12}$ after $P_3$, and the microscopic dephasing is inferred from the decay of the photon echo amplitude versus $\tau_{12}$. Conversely, the decay of the photon-echo amplitude versus $\tau_{23}$ probes the exciton density dynamics\,\cite{ShahBook96}.

The measured FWM field amplitude versus $\tau_{12}$ is given in \Fig{fig:Dephasing} by the detected voltage using an amplification of $10^5$\,V/A of the current from the silicon photodiodes in the balanced detection having a quantum efficiency of 0.75 and applying a reference power of about 0.3\,mW per diode. Measurements were taken at $\tau_{23}=1$\,ps to exclude non-resonant nonlinearities. The time-averaged excitation intensity was 17\,W/cm$^{2}$ per beam, within the third-order nonlinear response regime and resulting in negligible local heating, as we affirmed by power-dependent measurements. To minimize selective excitation of linearly polarized transitions in the ensemble of randomly oriented NPLs, all pulses were co-circularly polarized. The decay of the TI-FWM versus $\tau_{12}$ is described by two exponentially decaying components for temperatures $T$ above 10\,K, with an additional longer component visible for lower temperatures, as shown by fits to the data after pulse overlap $\tau_{12}>0.3\,$ps. The dynamics is somewhat dependent on the probed energy within the inhomogeneously broadened ensemble as shown in \Fig{fig:Dephasing}b. The FWHM homogeneous linewidths $2\hbar\gamma$ of the fitted dephasing rates $\gamma_1>\gamma_2$ (see inset) show that $\gamma_1$ slightly increases with increasing energy across the inhomogeneous distribution, together with its relative weight (shown by the ZPL weight as discussed later), while $\gamma_2$ is slightly decreasing.

\begin{figure}[t]
\centerline{\includegraphics*[width=8cm]{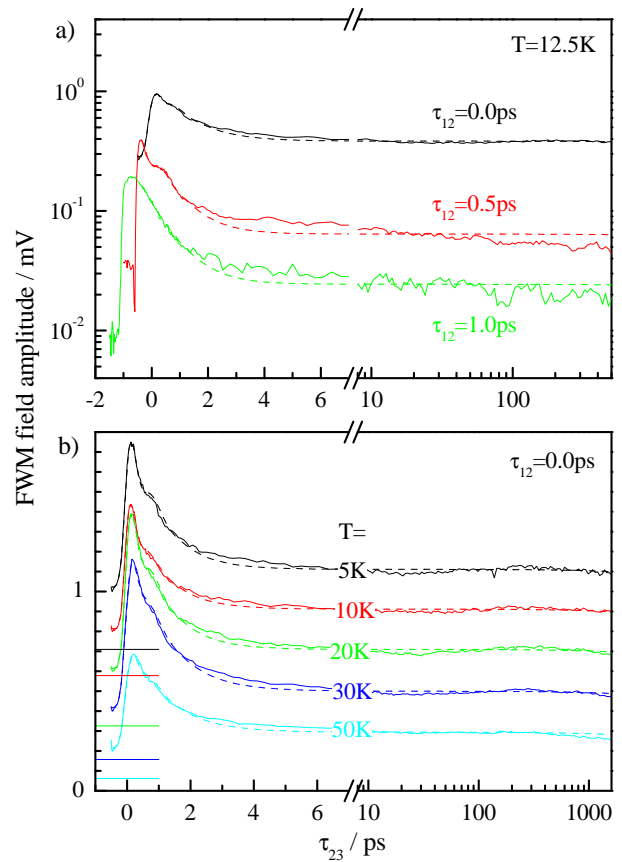}}
\caption{Exciton density dynamics measured from the TI-FWM field amplitude versus $\tau_{23}$ at fixed $\tau_{12}$. Dashed lines are fits to the data. a) Measurements at 12.5\,K for different values of $\tau_{12}$, as indicated. b) Measurements at $\tau_{12}=0$\,ps for different temperatures, as indicated.\label{fig:Density}}
\end{figure}

To investigate the physical origin of the observed dephasing, we have measured the exciton population dynamics by varying the delay time $\tau_{23}$, for different $\tau_{12}$ and temperatures. The measured response for $\tau_{12}=0$ shown in \Fig{fig:Density}b can be described by two exponential decays with weakly temperature dependent times around 1\,ps and 40\,ns. The latter is giving rise to a signal at $\tau_{23}<0$ due to a pile-up of the response from previous pulse repetitions of 13\,ns period which excited the sample earlier. All data were consistently fitted by the sum of two exponential decays including the pile-up effect yielding the decay rates $\Gamma_1>\Gamma_2$ and amplitudes $A_{1,2}$. Interestingly, when changing $\tau_{12}$ from 0 to 1\,ps, $A_1/A_2$ increases, such that the relevance of the pile-up effect decreases, while the rates are unchanged within error. This shows that the density induced absorption of the long lifetime component is spectrally broader than that of the the short one.

A possible origin of the $\Gamma_2$ component could be the spin-forbidden dark excitonic state. However, since we find that $\Gamma_2$ is nearly temperature independent from 5\,K to 50\,K, we can estimate the related dark-bright splitting $\deltaDB>k_{\rm B}T\log(\Gamma_2/\Gamma_1)\sim 40$\,meV for $T=50$\,K. This is much larger than the 1-10 meV found in colloidal CdSe QDs. We also do not find evidence for an internal relaxation between different bright/dark excitonic states, which modifies the dynamics for $\tau_{12}\neq0$ as observed on spherical nanocrystals\, \cite{MasiaPRL12,AccantoACSN12}. A more likely interpretation is charging of the NPL by carrier trapping in the surrounding, leaving a long-lived remaining carrier, which is also consistent with the spectral broadening of the response.

\begin{figure}[t]
\centerline{\includegraphics*[width=8cm]{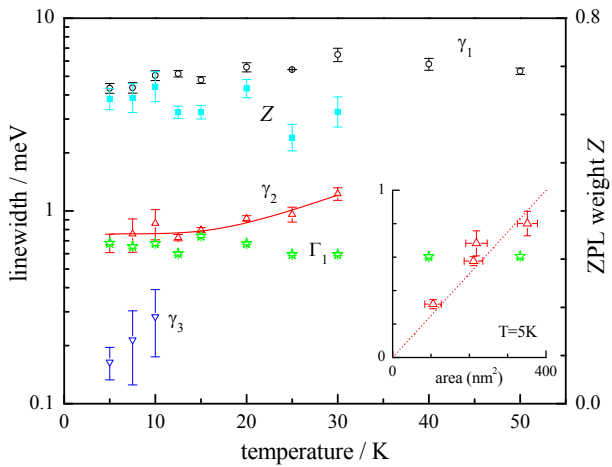}}
\caption{\label{fig:Rates} Linewidths and zero-phonon-line weight versus temperature. Shown are the dephasing linewidths $2\hbar\gamma_{1,2,3}$, and the lifetime limited linewidth $\hbar\Gamma_1$.  The line is a fit to the data for $\gamma_{2}$ . The ZPL weight $Z$ has been deduced from the amplitude of the slower components $\gamma_{1,2}$ relative to the maximum FWM signal calculated following the procedure in \Onlinecite{BorriPRB05}. The inset shows the measured $\gamma_2$ at T=5\,K as function of the NPL area.}
\end{figure}

The FWHM linewidth $\hbar\Gamma_1$ due to the density decay and the homogeneous width $2\hbar\gamma_{1,2}$ are shown in \Fig{fig:Rates} as a function of temperature. Remarkably, $2\gamma_{2}$ is equal to $\Gamma_1$ within error for $T\lesssim10$\,K. We therefore attribute $\gamma_2$ to the zero-phonon line (ZPL) dephasing of the BX transition in NPLs which is lifetime limited at low temperature. The deduced low-temperature ZPL width of $2\hbar\gamma_0=0.7$\,meV is consistent with PL linewidths measured on individual NPLs at low temperature \,\cite{TessierACSN12,AchtsteinNL12}, and about two orders of magnitude larger than in spherical QDs, where coherence times of up to 100\,ps, corresponding to 6\,$\mu$eV line widths, have been measured \cite{MasiaPRL12, AccantoACSN12}. The temperature dependence of $\gamma_2$ shown in \Fig{fig:Rates} is fitted by a temperature activated behavior $2\hbar\gamma_2=2\hbar\gamma_0+b/(\exp(\Delta/k_{\rm B}T)-1)$, yielding a spontaneous scattering rate $b=6$\,meV and an activation energy $\Delta=7\pm3$\,meV. Extrapolating to room temperature yields a homogeneous width of about 20\,meV, below the measured single NPL width of about 40\,meV \,\cite{TessierACSN12} which additionally contains scattering by LO phonons \cite{AchtsteinNL12}. The line-narrowing in the \Xhh\ absorption from room-temperature to low temperature seen in \Fig{fig:AbsTEM} is consistent with these values.
To discuss the scattering process leading to the dephasing, we have estimated the energy separation between the BX state and the first excited state from the quantization of the exciton CM motion. We use the "exciton-in-a-box" quantization energy $\frac{\hbar^2\pi^2}{2M}\left(\frac{n_x^2}{(L_x-2\aB)^2}+\frac{n_y^2}{(L_y-2\aB)^2}\right)$ where $n_{x,y}=1,2,...$ are the quantum numbers, $\aB$ is the in-plane exciton Bohr radius of about 2\,nm, and the exciton mass $M=0.37\me$ as sum of electron and hole mass from the Pigeon Brown model. The resulting energy separation of the BX $(n_x,n_y)=(1,1)$ to the first excited state $(1,2)$ is 4\,meV, which is similar to $\Delta$. The temperature dependence of $\gamma_2$ could thus be related to scattering into the (1,2) state by acoustic phonon absorption. Note that the (1,2) state has an odd parity and is thus dark.

The weak longer dephasing component $\gamma_3<\gamma_2$  shown in \Fig{fig:Rates} is attributed to a fraction of NPLs in the ensemble having a longer excitonic lifetime. This is compatible with the exciton dynamics shown in \Fig{fig:Density}, since a 13\% fraction of NPLs having a $\sim5$\,ps lifetime as given by relative amplitude $A_3/(A_3+A_2)$ results in an insignificant modification in the dynamics due to the strong $\Gamma_2$ component. When this component is  suppressed (see $\tau_{12}=1\,$ps in \Fig{fig:Density}a), a weak component with a decay time of about 5\,ps, consistent with the lifetime-limited density, is observed.

The dephasing rate $\gamma_1$ has a relative amplitude which increases with increasing temperature, indicating that this fast initial dephasing is containing phonon--assisted transitions. It is known that excitons confined in quantum dots exhibit a non-Lorentzian homogeneous lineshape, consisting of a sharp zero-phonon line superimposed onto a few meV wide acoustic phonon band which in turn gives rise to an initial fast dephasing\,\cite{BorriPRB05,MasiaPRL12}. Since excitons in the investigated NPLs are confined in a larger volume ($\sim 500$\,nm$^3$)  than in the nanocrystals studied in \Onlinecite{MasiaPRL12} ($\sim 200$\,nm$^3$), we expect a higher ZPL weight $Z$. We estimate $Z$ from the dephasing dynamics following the procedure discussed in \Onlinecite{BorriPRB05} using the sum amplitude associated with $\gamma_{2,3}$ relative to the maximum FWM signal around $\tau_{12}=0$. As shown in \Fig{fig:Rates}, we find values of $Z\sim0.6$ at low temperature, which is actually comparable to spherical nanocrystals \cite{MasiaPRL12}.  We attribute this smaller than expected $Z$ to an enhancement of the phonon-assisted transitions by the excited exciton states on the high energy side of the ZPL, leading to exciton-polaron transitions \cite{GladilinPRB04}. This attribution is supported by the observed decrease of $Z$ with increasing energy within the \Xhh\ absorption line (see \Fig{fig:Dephasing}b). Interestingly, single NPL spectra at $T=20$\,K \cite{TessierACSN12} show  an emission peak with a satellite shifted by about 4\,meV to higher energies having a relative weight of about 10\%. Considering that the  Boltzmann factor of thermal population for 4\,meV separation is about 0.1, we can estimate that this satellite has a similar absorption as the main peak, consistent with the ZPL weight of $Z=0.6$ deduced from the FWM dynamics. The energy separation of the exciton-polaron transitions is smaller than the pulse width of 16\,meV, such that we can excite both transitions simultaneously and we can see the resulting onset of a beat visible in the dynamics. The beat is strongly damped due to the large linewidth $\gamma_1$, which is consistent with the expected linewidth given by the spontaneous emission coefficient $b$ deduced from the temperature dependence of $\gamma_2$.

A related damped oscillation with a period of about 1\,ps is observed in the exciton density decay dynamics in \Fig{fig:Density}. These oscillations can be assigned to the modulation of the excitonic absorption by the coherent phonons created by the impulsive excitation, and have been previously observed in a variety of structures including  CdSe \cite{MittlemanPRB94, DworakPRL11} and PbS \cite{KraussPRL97} QDs. Interestingly, varying $\tau_{12}$ the visibility of the oscillations can be controlled \cite{MittlemanPRB94}. Specifically, for $\omp\tau_{12}=0,2\pi$ ($\tau_{12}\sim 0,1$\,ps), where $\omp$ is the angular phonon frequency, the oscillations are suppressed, while for $\omp\tau_{12}=\pi$ ($\tau_{12}\sim 0.5$\,ps) they are enhanced.
We have modeled the oscillations in the fit by multiplying the bi-exponential decay by $(1+B\exp(-\gp\tau_{12})\cos(\omp\tau_{12}))$ with the phonon decay rate $\gp$ and the amplitude $B$. The fit to the data (see \Fig{fig:Density}) yields $\hbar\omp=(4.1\pm 0.1)$\,meV and $\hbar\gp=(1.5 \pm 0.2)$\,meV. The oscillation frequency is similar to the expected frequency of the lowest longitudinal acoustic (LA) phonon mode confined by the NPL thickness given by $\hbar\omp=hv/(2L_z)=4.2$\,meV, with the LA velocity in CdSe $v=3.7\times10^3$\,m/s \cite{AdachiBook04}. The in-plane wavevector of the fundamental phonon mode excited by the delocalized exciton density can be neglected since $L_z\ll L_x,L_y$. The mode damping $\gp$ expected due to the transmission to the polystyrene environment can be estimated using the amplitude reflection coefficient of $r=0.75$ of the CdSe/PS interface calculated using the acoustic impedance mismatch, yielding $\hbar\gp=\hbar\ln(r)v/L_z=0.39$\,meV. This result is significantly smaller than the fitted value, which could be due to an inhomogeneous distribution of oscillation frequencies in the measured NPL ensemble, for example due to the varying coupling to PS considering that $L_z$ corresponds to only 6 styrene ring diameters.

Let us now discuss the physical interpretation of the measured exciton lifetime of about 1\,ps. When compared to the $\sim10$\,ns radiative lifetime in CdSe spherical nanocrystals\,\cite{CrookerAPL03} this lifetime is remarkably short. It is known that with increasing exciton COM extension the radiative lifetime decreases\,\cite{FeldmannPRL87,AndreaniPRB99}, an effect also referred to as "giant oscillator strength" \cite{FeldmannPRL87,AchtsteinNL12}. The radiative lifetime of a heavy-hole exciton has been calculated to be 12\,ps in a 10\,nm wide GaAs/AlGaAs QW\,\cite{AndreaniBook94}, and measured to be about 1\,ps in 10\,nm wide ZnSe/ZnMgSSe QWs \cite{LangbeinPSSA02a} and 16-20\,nm wide ZnSe/ZnMgSe QWs\, \cite{WagnerPRB98}. Considering the large exciton binding energy $\Rhh\sim180$\,meV  compared to the ZnSe QWs which have an exciton binding energy $\sim25$\,meV, we expect a free exciton radiative lifetime in extended NPLs in the order of 100\,fs. The measured lifetime of $\sim1$\,ps is thus consistent with excitons being localized in-plane by the lateral size of the NPLs, whereby the lifetime increases due to the reduced coherence area \,\cite{SavonaPRB06}. For NPL much smaller than the wavelength and much larger than \aB, the radiative rate is expected to be proportional to the NPL area. The measured low-temperature $\gamma_2$ for NPLs of different sizes $(24\times5,27\times8,31\times7,27\times13)\,$nm$^2$ is given in the inset of \Fig{fig:Rates}. We find  a proportionality to the NPL area, as expected for radiative decay. The weak component of rate $\gamma_3$ can be attributed to excitons with a smaller coherence area, possibly due to lateral disorder, leading to exciton localization within the NPL. Such lateral disorder can be due to NPL thickness variations or disorder in the dielectric surrounding. We note that the in-plane confinement energy created by a monolayer thickness variation is about 200\,meV, while the localization potential required to confine the exciton to an area which is a factor of $\gamma_2/\gamma_3$ smaller is about 50\,meV.

In conclusion, we have presented evidence of an intrinsic radiative lifetime in the 1\,ps range in quasi-2D CdSe nanoplatelets from dephasing and density dynamics measured by three-beam four-wave mixing. The radiative rate is scaling with the exciton coherence area, promising a tuning range from hundreds of picoseconds down to sub-picoseconds adjusting the platelet area, and merging the size tunability and monolayer thickness precision of colloidal synthesis with the large oscillator strength of quantum well excitons. Importantly, the nanoplatelets could be suited to reach the strong light-matter coupling regime in tuneable microcavities \cite{DolanOL10}, and enable applications as single photon switches.

\begin{acknowledgments}
A.N. acknowledges financial support by the President's Research Scholarship programme of Cardiff University, FM acknowledges financial support from the UK EPSRC Research Council (grant EP/H45848/1) and from the European Union (Marie Curie grant agreement PERG08- GA-2010-276807). P.B. acknowledges the UK EPSRC Research Council for her Leadership fellowship award (grant EP/I005072/1). S.C. and I.M. acknowledge funding from the European Union (REA grant agreement PIEF-GA-2011-298022). The authors thank Roberta Ruffilli for the preparation and measurement of the cross-section TEM samples.
\end{acknowledgments}

\section*{Supplement}
\subsection{NPL synthesis}
{\em Chemicals:} Cadmium nitrate tetrahydrate, cadmium acetate dihydrate Cd(Ac)$_2\cdot$2H$_2$O, technical grade 1-octadecene (ODE), oleic acid and sodium myristate were purchased from Sigma-Aldrich. Selenium (Se) mesh 99.99\% was purchased from STREM.

{\em Synthesis of cadmium myristate (Cd(myr)$_2$)}: 5\,g (0.02\,mol) of sodium myristate was dissolved in 150\,mL of methanol by stirring the solution for 1\,hour at room temperature. After addition of a solution of 3\,g of cadmium nitrate in 10\,mL of methanol, a white powder formed which was collected and dried for two days under vacuum.

{\em Synthesis of CdSe nanoplatelets emitting around 515\,nm}: 170\,mg of Cd(myr)$_2$ (0.3\,mmol), 12\,mg of Se and 15\,mL of ODE were added in a three-neck flask and degassed under vacuum. The mixture was heated under argon flow to 210$^\circ$C, and when this temperature was reached, 90\,mg of Cd(Ac)$_2$ were swiftly introduced. The mixture was further heated to 240$^\circ$C and kept at this temperature for 10 minutes. The NPL solution also contained a fraction of spherical quantum dots, which were separated from the NPLs by selective precipitation. After synthesis the average NPLs size and corresponding standard deviation were measured with transmission electron microscopy, evaluating the length and width of 70 particles. The X-ray diffraction pattern was measured on a drop-casted thin film of NPL using miscut silicon substrates.

\subsection{Colloid extinction and photoluminescence}

\begin{figure}[t]
\centerline{\includegraphics[width=8cm]{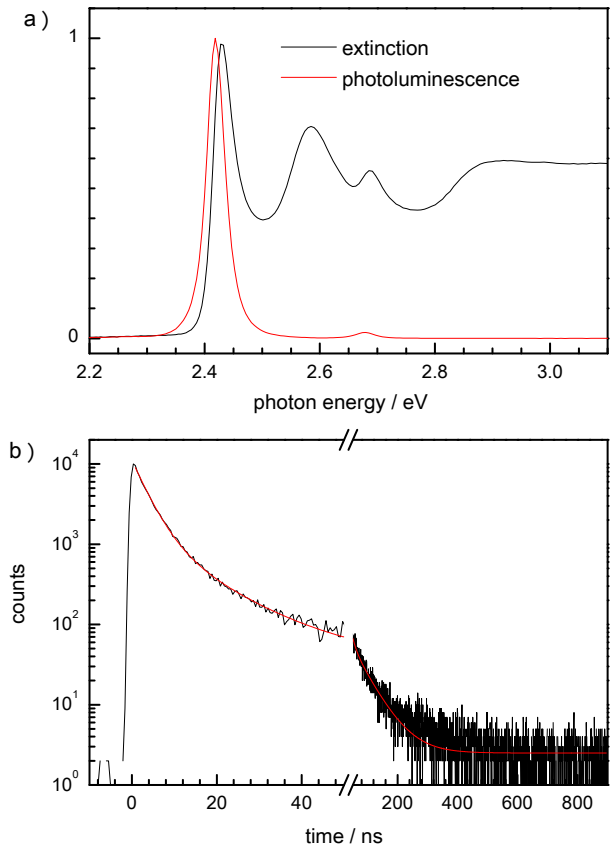}}
\caption{Optical properties of the investigated CdSe NPL colloid in toluene at room temperature. a) Normalized extinction (black line) and photoluminescence (red line) spectra. b) Time-resolved photoluminescence black line:data, red line: triple-exponential fit yielding the time constants
$\tau_1=3.36\pm0.06$\,ns, $\tau_2=11.2\pm0.4$\,ns, $\tau_3=61\pm 2$\,ns and amplitudes
$A_1=9540\pm110$, $A_2=1530\pm111$, $A_3=112\pm6$. \label{fig:solutionAbsPL}}
\end{figure}

The optical properties of the NPL colloid used to prepare the samples were characterized. Photoluminescence (PL), excited at 400\,nm wavelength, and extinction spectra are shown in \Fig{fig:solutionAbsPL}b. The colloid is dominated by 6ML NPLs, but also 5ML thick NPLs are visible, having about 5\% number fraction as estimated from the heavy hole exciton extinction peak. The PL is dominated by excitonic emission with a Stokes shift of about 11\,meV, and a linewidth of 38\,meV. The time-resolved PL is shown in \Fig{fig:solutionAbsPL}b and was excited with 50\,ps pulses at 400\,nm with 1\,MHz repetition rate and detected using time-correlated single photon counting with a time-resolution of 0.7\,ns. It reveals an initial decay time of about 3\,ns accounting for 57\% of the total emission, followed by a component with 11\,ns accounting for 30\%, and 60\,ns accounting for 12\%.

This dynamics is comparable to reports of the PL dynamics in literature \cite{TessierACSN12}. When a higher time-resolution in employed, also components of faster emission dynamics in the 100\,ps range have been reported \cite{AchtsteinNL12}.

\subsection{Low-temperature extinction spectra}
Low-temperature extinction spectra were measured with a tungsten white-light source and a Ocean Optics HR4000 spectrometer. The transmitted spectral intensity $I_{\rm s}(\omega)$ through the sample was measured over a sample region of 20\,$\mu$m diameter, and compared with a reference transmission $I_{\rm r}(\omega)$ at a position without polymer film laterally offset from the sample of about 0.5\,mm lateral size and 10\,$\mu$m thickness. The resulting extinction is given by $\aext(\omega)=\ln(I_{s}/I_{\rm r})$ and is shown in \Fig{fig:AbsorptionFit}a. It resembles the room temperature extinction measured in the colloid shown in \Fig{fig:solutionAbsPL}a, but is shifted to higher energies due to the temperature dependent band-gap shift. Furthermore, we observe a reduced linewidth of the heavy-hole exciton, attributed to reduced phonon-scattering, similar to other reports in literature \cite{AchtsteinNL12}.

\subsection{Extinction fit}
The extinction of the NPL ensemble was fitted using a quantum-well absorption model, consisting for each interband transition of the 1s exciton absorption and a continuum edge, with a line-shape
\be p(\omega)=\pX+\frac{L/\pi}{\gL+\Delta^2/\gL}+\frac{\AC}{2}\left[1+\erf\left(\frac{\Delta-\omB}{\gC} \right)\right]\label{eq:AbsorptionFit}\ee
where $\Delta=\omega-\omega_0$, with the exciton energy $\omega_0$, the line-width $\gamma$, and the exciton binding energy $\omB$. The continuum step height is $A_{\rm C}$ and the step width is $\gC$. The 1s exciton line-shape is modeled by an absorption shape $\pX$ of unity area and an additional broader Lorentzian of width $\gL$ and area $L$, accounting for the observed tails attributed to Rayleigh scattering.

\begin{figure}[t]
\centerline{\includegraphics[width=8cm]{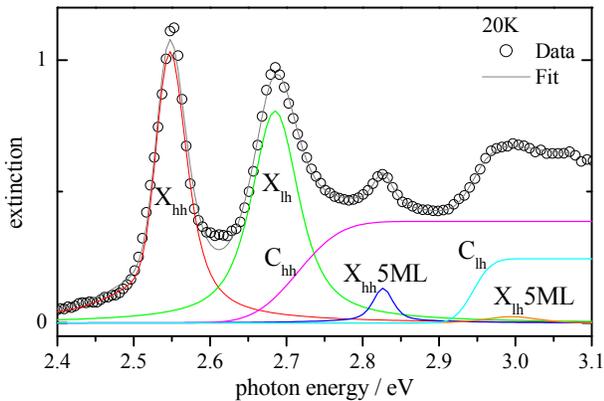}}
\caption{Measured NPL extinction spectrum (circles) and fit (black line) with its individual components as labeled. {\bf WL: Ali, add missing parameters}\label{fig:AbsorptionFit}}
\end{figure}

The absorption line-shape of a quantum well exciton can be modeled accurately by an asymmetric lineshape taking into account the in-plane localization \cite{SchnabelPRB92, LeossonPRB00}
\be \pX=\frac{1}{2\eta}\left[1+\erf\left(\frac{\Delta}{\gamma}-\frac{\gamma}{2\eta}\right)\right] \exp\left(\frac{\gamma^2}{4\eta^2}-\frac{\Delta}{\eta}\right) \label{eqn:peakassym}\ee
with the additional parameter $\eta$ describing the asymmetric broadening by localization. In case the inhomogeneous broadening is not significant, we use the simpler function
\be \pX=\frac{1}{2\gamma\cosh^2\left(\Delta^2/\gamma^2\right)} \label{eqn:peakcosh}\ee
with a FWHM of $1.763\,\gamma$, having exponential tails, which was giving better fits than a Gaussian line-shape. For the heavy-hole band we used $p_{\rm hh}$ according to \Eq{eqn:peakassym}. For the light-hole band instead, we used $p_{\rm lh}$ given by the simpler \Eq{eqn:peakcosh} since the exciton shows a larger homogeneous broadening due to scattering into the overlapping heavy-hole exciton continuum, as known from quantum wells \cite{PasquarelloPRB91}. The resulting lineshape is
\be \aext^{\rm 6ML}=A_{\rm hh}p_{\rm hh}+A_{\rm lh}p_{\rm lh}\ee
with the weights $A_{\rm hh}$ and $A_{\rm lh}$ of the bands. Since the sample shows a small amount of 5ML platelets in both PL and extinction, we have fitted a 5ML extinction spectrum at room temperature separately, and added this fit $\aext^{\rm 5ML}$,  with an amplitude factor and a rigid energy shift due to the temperature difference, to arrive at $\aext=\aext^{\rm 6ML}+A_{\rm 5ML}\aext^{\rm 5ML}$.
The parameters $L$ and $\gamma_L$ were chosen equal for all bands. The resulting fit to the NPL extinction data at 20K is shown in \Fig{fig:AbsorptionFit}, showing also the individual components. The corresponding fit parameters with fit error estimates are summarized in \Tab{tab:AbsorptionFit}.

\begin{table}
\begin{tabular}{r|c|c|c|c|}
 & hh 6ML & lh 6ML& hh 5ML & lh 5ML\\
\hline
$L$ & \multicolumn{4}{c|}{$2.1 \pm 0.2$}\\
$\hbar\gL$/meV & \multicolumn{4}{c|}{$0.145 \pm 0.017$}\\ \cline{2-5}
$A\times 10^2$meV & $4.16\pm0.2$ & $4.37\pm0.95$ & $0.38\pm0.09$ & $0.22\pm0.05$ \\
$\hbar\omega_0/$meV &  $2536.9 \pm 1.5$ & $2685.2\pm 1.6$ & $2826.5\pm 3$ & $2826.5\pm 3$\\
$\hbar\omB/$meV & $178 \pm 34$ & $259 \pm 3$ & - & -\\
$\hbar\gamma/$meV & $21.2 \pm 0.9$ & $36.1 \pm 2.0$ & $16.6 \pm 4$ & $46.1\pm11$\\
$\hbar\eta/$meV & $16.3 \pm 2.5 $ & - & - & -\\
$\hbar\gC/$meV & $63 \pm 30$ & $27 \pm 4$ & - & -\\
$\AC\times$meV & $9.3 \pm 0.5$ & $5.6 \pm 1.2$ & - & -\\
\hline
\end{tabular}
\caption{Fit parameters with errors of the absorption fit shown in \Fig{fig:AbsorptionFit}, yielding an $R^2=0.99427$. \label{tab:AbsorptionFit}}
\end{table}

It is worth noting that zinc-blende CdSe has an additional interband transition due to the split-off valence band with a separation of about $\Delta_{\rm so}=0.39$\,eV to the hh and lh band, which results in an additional excitonic transition. Its energy should be shifted by $\Delta_{\rm so}$ and by the quantization energy which is expected to be in-between the one of heavy hole and light hole band according to the effective mass being in-between the ones of these bands. We therefore would expect an additional absorption around 3\,eV from the split-off 1s exciton, which is close to the fitted position of \Clh. This excitonic transition is expected to be significantly broadened by the decay into heavy-hole and light-hole excitons and continua, beyond what was observed for the light-hole exciton.

Changing the NPL thickness, the ratio between $\Delta_{\rm so}$ and the exciton binding energies changes. Indeed, the extinction spectra shown in \Onlinecite{IthurriaNMa11} for NPL of different thickness show a changing shape around $\Clh$. The 7ML NPL has a peak at the expected \Xso\ position, with $\Clh$ being below \Xso, while the 5ML NPL shows a double step structure with $\Clh$ being above \Xso. In the 6ML NPL, \Xso\ seems to be be just below  $\Clh$, resulting in a flat-top structure. The effect of the split-off exciton therefore might shift the apparent continuum edge to somewhat lower energies, leading to an underestimation of the lh exciton binding energy in the fit.
\subsection{Sample preparation in polymer}
The samples were prepared in the following way. A NPL toluene colloid with $20\mu$M concentration was mixed with a solution of $5\%$ weight polystyrene of average molecular weight 280000 in toluene. The volume ratio of NPL colloid to polystyrene solution was $8:2$. The resulting solution was drop-cast onto a microscope slide placed on a hotplate at a temperature of $80^\circ$C. Four layers of $5\mu$L dropcasts were made on top of each other. Each dropcast was allowed to dry for a few minutes before the next one was applied. A suitable region of the resulting film was cut to a size of approximately $0.5\times0.5$\,mm$^2$ and squeezed between two 1\,mm thick quartz windows of 5\,mm diameter using a home-made sample holder. The sample holder was then placed on a hot plate at $120^\circ$C for a few seconds and the two quartz windows were further squeezed together to ensure good thermal contact with the sample.

NPL aggregation in the sample could change their dynamics\cite{TessierACSN13} due to F{\"o}rster-type interactions, typically relevant for distances below 10\,nm.
To verify that this is not a significant effect in the investigated samples, we have measured the spatial distribution of the NPL in the polymer sample by transmission electron microscopy (TEM). Tiny film fragments were embedded in a super glue drop on a polymeric support, in order to handle the samples to be cut for the cross sectional studies. Sections of about 70 nm were cut with a diamond knife (Diatome) on a Leica EM UC6 ultramicrotome. TEM images were collected by a FEI Tecnai G2 F20 equipped with a field-emission gun (FEG), operating at 200 kV of acceleration voltage and recorded with a 4 Mp Gatan BM UltraScan Charge-Coupled Device (CCD) camera.

\begin{figure}[t]
\centerline{\includegraphics[width=\columnwidth]{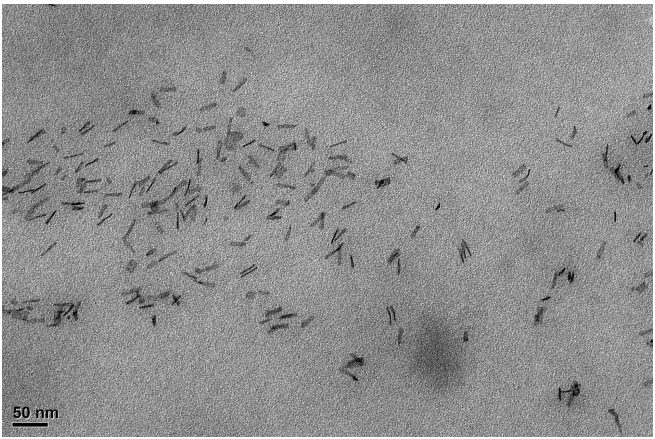}}
\caption{Transmission electron microscopic image of a sample of CdSe NPL in PS prepared in the same way as the sample studied in the dephasing measurements. The scale bar is shown\label{fig:TEMPolymer}}
\end{figure}

A representative high resolution image of a section of the sample is given in \Fig{fig:TEMPolymer}. A small fraction shows distances below 10\,nm. We do not expect that this type of aggregation significantly influences the dynamics. Lower resolution images reveals some bunching on a micrometer length scale. This might be the origin of the stronger Rayleigh scattering tails observed in the extinction spectra compared to solution measurements.
%


\begin{thebibliography}{33}%
\makeatletter
\providecommand \@ifxundefined [1]{%
 \@ifx{#1\undefined}
}%
\providecommand \@ifnum [1]{%
 \ifnum #1\expandafter \@firstoftwo
 \else \expandafter \@secondoftwo
 \fi
}%
\providecommand \@ifx [1]{%
 \ifx #1\expandafter \@firstoftwo
 \else \expandafter \@secondoftwo
 \fi
}%
\providecommand \natexlab [1]{#1}%
\providecommand \enquote  [1]{``#1''}%
\providecommand \bibnamefont  [1]{#1}%
\providecommand \bibfnamefont [1]{#1}%
\providecommand \citenamefont [1]{#1}%
\providecommand \href@noop [0]{\@secondoftwo}%
\providecommand \href [0]{\begingroup \@sanitize@url \@href}%
\providecommand \@href[1]{\@@startlink{#1}\@@href}%
\providecommand \@@href[1]{\endgroup#1\@@endlink}%
\providecommand \@sanitize@url [0]{\catcode `\\12\catcode `\$12\catcode
  `\&12\catcode `\#12\catcode `\^12\catcode `\_12\catcode `\%12\relax}%
\providecommand \@@startlink[1]{}%
\providecommand \@@endlink[0]{}%
\providecommand \url  [0]{\begingroup\@sanitize@url \@url }%
\providecommand \@url [1]{\endgroup\@href {#1}{\urlprefix }}%
\providecommand \urlprefix  [0]{URL }%
\providecommand \Eprint [0]{\href }%
\providecommand \doibase [0]{http://dx.doi.org/}%
\providecommand \selectlanguage [0]{\@gobble}%
\providecommand \bibinfo  [0]{\@secondoftwo}%
\providecommand \bibfield  [0]{\@secondoftwo}%
\providecommand \translation [1]{[#1]}%
\providecommand \BibitemOpen [0]{}%
\providecommand \bibitemStop [0]{}%
\providecommand \bibitemNoStop [0]{.\EOS\space}%
\providecommand \EOS [0]{\spacefactor3000\relax}%
\providecommand \BibitemShut  [1]{\csname bibitem#1\endcsname}%
\let\auto@bib@innerbib\@empty
\bibitem [{\citenamefont {Ithurria}\ and\ \citenamefont
  {Dubertret}(2008)}]{IthurriaJACS08}%
  \BibitemOpen
  \bibfield  {author} {\bibinfo {author} {\bibfnamefont {S.}~\bibnamefont
  {Ithurria}}\ and\ \bibinfo {author} {\bibfnamefont {B.}~\bibnamefont
  {Dubertret}},\ }\href@noop {} {\bibfield  {journal} {\bibinfo  {journal} {J.
  Am. Chem. Soc.}\ }\textbf {\bibinfo {volume} {130}},\ \bibinfo {pages}
  {16504} (\bibinfo {year} {2008})}\BibitemShut {NoStop}%
\bibitem [{\citenamefont {Ithurria}\ \emph {et~al.}(2011)\citenamefont
  {Ithurria}, \citenamefont {Tessier}, \citenamefont {Mahler}, \citenamefont
  {Lobo}, \citenamefont {Dubertret},\ and\ \citenamefont
  {Efros}}]{IthurriaNMa11}%
  \BibitemOpen
  \bibfield  {author} {\bibinfo {author} {\bibfnamefont {S.}~\bibnamefont
  {Ithurria}}, \bibinfo {author} {\bibfnamefont {M.~D.}\ \bibnamefont
  {Tessier}}, \bibinfo {author} {\bibfnamefont {B.}~\bibnamefont {Mahler}},
  \bibinfo {author} {\bibfnamefont {R.~P. S.~M.}\ \bibnamefont {Lobo}},
  \bibinfo {author} {\bibfnamefont {B.}~\bibnamefont {Dubertret}}, \ and\
  \bibinfo {author} {\bibfnamefont {A.~L.}\ \bibnamefont {Efros}},\ }\href
  {\doibase 10.1038/NMAT3145} {\bibfield  {journal} {\bibinfo  {journal} {Nat.
  Mater.}\ }\textbf {\bibinfo {volume} {10}},\ \bibinfo {pages} {936} (\bibinfo
  {year} {2011})}\BibitemShut {NoStop}%
\bibitem [{\citenamefont {Tessier}\ \emph {et~al.}(2012)\citenamefont
  {Tessier}, \citenamefont {Javaux}, \citenamefont {Maksimovic}, \citenamefont
  {Loriette},\ and\ \citenamefont {Dubertret}}]{TessierACSN12}%
  \BibitemOpen
  \bibfield  {author} {\bibinfo {author} {\bibfnamefont {M.~D.}\ \bibnamefont
  {Tessier}}, \bibinfo {author} {\bibfnamefont {C.}~\bibnamefont {Javaux}},
  \bibinfo {author} {\bibfnamefont {I.}~\bibnamefont {Maksimovic}}, \bibinfo
  {author} {\bibfnamefont {V.}~\bibnamefont {Loriette}}, \ and\ \bibinfo
  {author} {\bibfnamefont {B.}~\bibnamefont {Dubertret}},\ }\href {\doibase
  10.1021/nn3014855} {\bibfield  {journal} {\bibinfo  {journal} {ACS Nano}\
  }\textbf {\bibinfo {volume} {6}},\ \bibinfo {pages} {6751} (\bibinfo {year}
  {2012})}\BibitemShut {NoStop}%
\bibitem [{\citenamefont {Leosson}\ \emph {et~al.}(2000)\citenamefont
  {Leosson}, \citenamefont {Jensen}, \citenamefont {Langbein},\ and\
  \citenamefont {Hvam}}]{LeossonPRB00}%
  \BibitemOpen
  \bibfield  {author} {\bibinfo {author} {\bibfnamefont {K.}~\bibnamefont
  {Leosson}}, \bibinfo {author} {\bibfnamefont {J.~R.}\ \bibnamefont {Jensen}},
  \bibinfo {author} {\bibfnamefont {W.}~\bibnamefont {Langbein}}, \ and\
  \bibinfo {author} {\bibfnamefont {J.~M.}\ \bibnamefont {Hvam}},\ }\href@noop
  {} {\bibfield  {journal} {\bibinfo  {journal} {Phys. Rev. B}\ }\textbf
  {\bibinfo {volume} {61}},\ \bibinfo {pages} {10322} (\bibinfo {year}
  {2000})}\BibitemShut {NoStop}%
\bibitem [{\citenamefont {Voigt}\ \emph {et~al.}(1979)\citenamefont {Voigt},
  \citenamefont {Spiegelberg},\ and\ \citenamefont {Senoner}}]{VoigtPSSB79}%
  \BibitemOpen
  \bibfield  {author} {\bibinfo {author} {\bibfnamefont {J.}~\bibnamefont
  {Voigt}}, \bibinfo {author} {\bibfnamefont {F.}~\bibnamefont {Spiegelberg}},
  \ and\ \bibinfo {author} {\bibfnamefont {M.}~\bibnamefont {Senoner}},\ }\href
  {\doibase 10.1002/pssb.2220910120} {\bibfield  {journal} {\bibinfo  {journal}
  {physica status solidi (b)}\ }\textbf {\bibinfo {volume} {91}},\ \bibinfo
  {pages} {189} (\bibinfo {year} {1979})}\BibitemShut {NoStop}%
\bibitem [{\citenamefont {Muljarov}\ \emph {et~al.}(1995)\citenamefont
  {Muljarov}, \citenamefont {Tikhodeev}, \citenamefont {Gippius},\ and\
  \citenamefont {Ishihara}}]{MuljarovPRB95}%
  \BibitemOpen
  \bibfield  {author} {\bibinfo {author} {\bibfnamefont {E.~A.}\ \bibnamefont
  {Muljarov}}, \bibinfo {author} {\bibfnamefont {S.~G.}\ \bibnamefont
  {Tikhodeev}}, \bibinfo {author} {\bibfnamefont {N.~A.}\ \bibnamefont
  {Gippius}}, \ and\ \bibinfo {author} {\bibfnamefont {T.}~\bibnamefont
  {Ishihara}},\ }\href {\doibase 10.1103/PhysRevB.51.14370} {\bibfield
  {journal} {\bibinfo  {journal} {Phys. Rev. B}\ }\textbf {\bibinfo {volume}
  {51}},\ \bibinfo {pages} {14370} (\bibinfo {year} {1995})}\BibitemShut
  {NoStop}%
\bibitem [{\citenamefont {Achtstein}\ \emph {et~al.}(2012)\citenamefont
  {Achtstein}, \citenamefont {Schliwa}, \citenamefont {Prudnikau},
  \citenamefont {Hardzei}, \citenamefont {Artemyev}, \citenamefont {Thomsen},\
  and\ \citenamefont {Woggon}}]{AchtsteinNL12}%
  \BibitemOpen
  \bibfield  {author} {\bibinfo {author} {\bibfnamefont {A.}~\bibnamefont
  {Achtstein}}, \bibinfo {author} {\bibfnamefont {A.}~\bibnamefont {Schliwa}},
  \bibinfo {author} {\bibfnamefont {A.}~\bibnamefont {Prudnikau}}, \bibinfo
  {author} {\bibfnamefont {M.}~\bibnamefont {Hardzei}}, \bibinfo {author}
  {\bibfnamefont {M.}~\bibnamefont {Artemyev}}, \bibinfo {author}
  {\bibfnamefont {C.}~\bibnamefont {Thomsen}}, \ and\ \bibinfo {author}
  {\bibfnamefont {U.}~\bibnamefont {Woggon}},\ }\href@noop {} {\bibfield
  {journal} {\bibinfo  {journal} {Nano Lett.}\ }\textbf {\bibinfo {volume}
  {12}},\ \bibinfo {pages} {3151} (\bibinfo {year} {2012})}\BibitemShut
  {NoStop}%
\bibitem [{\citenamefont {Feldmann}\ \emph {et~al.}(1987)\citenamefont
  {Feldmann}, \citenamefont {Peter}, \citenamefont {G{\"o}bel}, \citenamefont
  {Dawson}, \citenamefont {Moore}, \citenamefont {Foxon},\ and\ \citenamefont
  {Elliott}}]{FeldmannPRL87}%
  \BibitemOpen
  \bibfield  {author} {\bibinfo {author} {\bibfnamefont {J.}~\bibnamefont
  {Feldmann}}, \bibinfo {author} {\bibfnamefont {G.}~\bibnamefont {Peter}},
  \bibinfo {author} {\bibfnamefont {E.~O.}\ \bibnamefont {G{\"o}bel}}, \bibinfo
  {author} {\bibfnamefont {P.}~\bibnamefont {Dawson}}, \bibinfo {author}
  {\bibfnamefont {K.}~\bibnamefont {Moore}}, \bibinfo {author} {\bibfnamefont
  {C.}~\bibnamefont {Foxon}}, \ and\ \bibinfo {author} {\bibfnamefont {R.~J.}\
  \bibnamefont {Elliott}},\ }\href@noop {} {\bibfield  {journal} {\bibinfo
  {journal} {Phys. Rev. Lett.}\ }\textbf {\bibinfo {volume} {59}},\ \bibinfo
  {pages} {2337} (\bibinfo {year} {1987})}\BibitemShut {NoStop}%
\bibitem [{\citenamefont {Andreani}\ \emph {et~al.}(1999)\citenamefont
  {Andreani}, \citenamefont {Panzarini},\ and\ \citenamefont
  {G{\'e}rard}}]{AndreaniPRB99}%
  \BibitemOpen
  \bibfield  {author} {\bibinfo {author} {\bibfnamefont {L.~C.}\ \bibnamefont
  {Andreani}}, \bibinfo {author} {\bibfnamefont {G.}~\bibnamefont {Panzarini}},
  \ and\ \bibinfo {author} {\bibfnamefont {J.-M.}\ \bibnamefont {G{\'e}rard}},\
  }\href@noop {} {\bibfield  {journal} {\bibinfo  {journal} {Phys. Rev. B}\
  }\textbf {\bibinfo {volume} {60}},\ \bibinfo {pages} {13276} (\bibinfo {year}
  {1999})}\BibitemShut {NoStop}%
\bibitem [{\citenamefont {Hours}\ \emph {et~al.}(2005)\citenamefont {Hours},
  \citenamefont {Senellart}, \citenamefont {Peter}, \citenamefont {Cavanna},\
  and\ \citenamefont {Bloch}}]{HoursPRB05}%
  \BibitemOpen
  \bibfield  {author} {\bibinfo {author} {\bibfnamefont {J.}~\bibnamefont
  {Hours}}, \bibinfo {author} {\bibfnamefont {P.}~\bibnamefont {Senellart}},
  \bibinfo {author} {\bibfnamefont {E.}~\bibnamefont {Peter}}, \bibinfo
  {author} {\bibfnamefont {A.}~\bibnamefont {Cavanna}}, \ and\ \bibinfo
  {author} {\bibfnamefont {J.}~\bibnamefont {Bloch}},\ }\href@noop {}
  {\bibfield  {journal} {\bibinfo  {journal} {Phys. Rev. B}\ }\textbf {\bibinfo
  {volume} {71}},\ \bibinfo {pages} {161306(R)} (\bibinfo {year}
  {2005})}\BibitemShut {NoStop}%
\bibitem [{\citenamefont {Savona}\ and\ \citenamefont
  {Langbein}(2006)}]{SavonaPRB06}%
  \BibitemOpen
  \bibfield  {author} {\bibinfo {author} {\bibfnamefont {V.}~\bibnamefont
  {Savona}}\ and\ \bibinfo {author} {\bibfnamefont {W.}~\bibnamefont
  {Langbein}},\ }\href@noop {} {\bibfield  {journal} {\bibinfo  {journal}
  {Phys. Rev. B}\ }\textbf {\bibinfo {volume} {74}},\ \bibinfo {pages} {075311}
  (\bibinfo {year} {2006})}\BibitemShut {NoStop}%
\bibitem [{\citenamefont {Borri}\ \emph {et~al.}(1999)\citenamefont {Borri},
  \citenamefont {Langbein}, \citenamefont {Hvam},\ and\ \citenamefont
  {Martelli}}]{BorriPRB99}%
  \BibitemOpen
  \bibfield  {author} {\bibinfo {author} {\bibfnamefont {P.}~\bibnamefont
  {Borri}}, \bibinfo {author} {\bibfnamefont {W.}~\bibnamefont {Langbein}},
  \bibinfo {author} {\bibfnamefont {J.~M.}\ \bibnamefont {Hvam}}, \ and\
  \bibinfo {author} {\bibfnamefont {F.}~\bibnamefont {Martelli}},\ }\href@noop
  {} {\bibfield  {journal} {\bibinfo  {journal} {Phys. Rev. B}\ }\textbf
  {\bibinfo {volume} {60}},\ \bibinfo {pages} {4505} (\bibinfo {year}
  {1999})}\BibitemShut {NoStop}%
\bibitem [{\citenamefont {Langbein}\ and\ \citenamefont
  {Hvam}(2000)}]{LangbeinPRB00}%
  \BibitemOpen
  \bibfield  {author} {\bibinfo {author} {\bibfnamefont {W.}~\bibnamefont
  {Langbein}}\ and\ \bibinfo {author} {\bibfnamefont {J.~M.}\ \bibnamefont
  {Hvam}},\ }\href@noop {} {\bibfield  {journal} {\bibinfo  {journal} {Phys.
  Rev. B}\ }\textbf {\bibinfo {volume} {61}},\ \bibinfo {pages} {1692}
  (\bibinfo {year} {2000})}\BibitemShut {NoStop}%
\bibitem [{\citenamefont {Borri}\ \emph {et~al.}(2001)\citenamefont {Borri},
  \citenamefont {Langbein}, \citenamefont {Schneider}, \citenamefont {Woggon},
  \citenamefont {Sellin}, \citenamefont {Ouyang},\ and\ \citenamefont
  {Bimberg}}]{BorriPRL01}%
  \BibitemOpen
  \bibfield  {author} {\bibinfo {author} {\bibfnamefont {P.}~\bibnamefont
  {Borri}}, \bibinfo {author} {\bibfnamefont {W.}~\bibnamefont {Langbein}},
  \bibinfo {author} {\bibfnamefont {S.}~\bibnamefont {Schneider}}, \bibinfo
  {author} {\bibfnamefont {U.}~\bibnamefont {Woggon}}, \bibinfo {author}
  {\bibfnamefont {R.~L.}\ \bibnamefont {Sellin}}, \bibinfo {author}
  {\bibfnamefont {D.}~\bibnamefont {Ouyang}}, \ and\ \bibinfo {author}
  {\bibfnamefont {D.}~\bibnamefont {Bimberg}},\ }\href@noop {} {\bibfield
  {journal} {\bibinfo  {journal} {Phys. Rev. Lett.}\ }\textbf {\bibinfo
  {volume} {87}},\ \bibinfo {pages} {157401} (\bibinfo {year}
  {2001})}\BibitemShut {NoStop}%
\bibitem [{\citenamefont {Borri}\ \emph {et~al.}(2005)\citenamefont {Borri},
  \citenamefont {Langbein}, \citenamefont {Woggon}, \citenamefont {Stavarache},
  \citenamefont {Reuter},\ and\ \citenamefont {Wieck}}]{BorriPRB05}%
  \BibitemOpen
  \bibfield  {author} {\bibinfo {author} {\bibfnamefont {P.}~\bibnamefont
  {Borri}}, \bibinfo {author} {\bibfnamefont {W.}~\bibnamefont {Langbein}},
  \bibinfo {author} {\bibfnamefont {U.}~\bibnamefont {Woggon}}, \bibinfo
  {author} {\bibfnamefont {V.}~\bibnamefont {Stavarache}}, \bibinfo {author}
  {\bibfnamefont {D.}~\bibnamefont {Reuter}}, \ and\ \bibinfo {author}
  {\bibfnamefont {A.~D.}\ \bibnamefont {Wieck}},\ }\href@noop {} {\bibfield
  {journal} {\bibinfo  {journal} {Phys. Rev. B}\ }\textbf {\bibinfo {volume}
  {71}},\ \bibinfo {pages} {115328} (\bibinfo {year} {2005})}\BibitemShut
  {NoStop}%
\bibitem [{\citenamefont {Masia}\ \emph {et~al.}(2011)\citenamefont {Masia},
  \citenamefont {Langbein}, \citenamefont {Moreels}, \citenamefont {Hens},\
  and\ \citenamefont {Borri}}]{MasiaPRB11}%
  \BibitemOpen
  \bibfield  {author} {\bibinfo {author} {\bibfnamefont {F.}~\bibnamefont
  {Masia}}, \bibinfo {author} {\bibfnamefont {W.}~\bibnamefont {Langbein}},
  \bibinfo {author} {\bibfnamefont {I.}~\bibnamefont {Moreels}}, \bibinfo
  {author} {\bibfnamefont {Z.}~\bibnamefont {Hens}}, \ and\ \bibinfo {author}
  {\bibfnamefont {P.}~\bibnamefont {Borri}},\ }\href@noop {} {\bibfield
  {journal} {\bibinfo  {journal} {Phys. Rev. B.}\ }\textbf {\bibinfo {volume}
  {83}},\ \bibinfo {pages} {201309(R)} (\bibinfo {year} {2011})}\BibitemShut
  {NoStop}%
\bibitem [{\citenamefont {Masia}\ \emph {et~al.}(2012)\citenamefont {Masia},
  \citenamefont {Accanto}, \citenamefont {Langbein},\ and\ \citenamefont
  {Borri}}]{MasiaPRL12}%
  \BibitemOpen
  \bibfield  {author} {\bibinfo {author} {\bibfnamefont {F.}~\bibnamefont
  {Masia}}, \bibinfo {author} {\bibfnamefont {N.}~\bibnamefont {Accanto}},
  \bibinfo {author} {\bibfnamefont {W.}~\bibnamefont {Langbein}}, \ and\
  \bibinfo {author} {\bibfnamefont {P.}~\bibnamefont {Borri}},\ }\href@noop {}
  {\bibfield  {journal} {\bibinfo  {journal} {Phys. Rev. Lett.}\ }\textbf
  {\bibinfo {volume} {108}},\ \bibinfo {pages} {087401} (\bibinfo {year}
  {2012})}\BibitemShut {NoStop}%
\bibitem [{\citenamefont {Accanto}\ \emph {et~al.}(2012)\citenamefont
  {Accanto}, \citenamefont {Masia}, \citenamefont {Moreels}, \citenamefont
  {Hens}, \citenamefont {Langbein},\ and\ \citenamefont
  {Borri}}]{AccantoACSN12}%
  \BibitemOpen
  \bibfield  {author} {\bibinfo {author} {\bibfnamefont {N.}~\bibnamefont
  {Accanto}}, \bibinfo {author} {\bibfnamefont {F.}~\bibnamefont {Masia}},
  \bibinfo {author} {\bibfnamefont {I.}~\bibnamefont {Moreels}}, \bibinfo
  {author} {\bibfnamefont {Z.}~\bibnamefont {Hens}}, \bibinfo {author}
  {\bibfnamefont {W.}~\bibnamefont {Langbein}}, \ and\ \bibinfo {author}
  {\bibfnamefont {P.}~\bibnamefont {Borri}},\ }\href {\doibase
  10.1021/nn300992a} {\bibfield  {journal} {\bibinfo  {journal} {ACS Nano}\
  }\textbf {\bibinfo {volume} {6}},\ \bibinfo {pages} {5227} (\bibinfo {year}
  {2012})}\BibitemShut {NoStop}%
\bibitem [{\citenamefont {Borri}\ and\ \citenamefont
  {Langbein}(2007)}]{BorriJPCM07}%
  \BibitemOpen
  \bibfield  {author} {\bibinfo {author} {\bibfnamefont {P.}~\bibnamefont
  {Borri}}\ and\ \bibinfo {author} {\bibfnamefont {W.}~\bibnamefont
  {Langbein}},\ }\href@noop {} {\bibfield  {journal} {\bibinfo  {journal} {J.
  Phys.: Condens. Matter.}\ }\textbf {\bibinfo {volume} {19}},\ \bibinfo
  {pages} {295201} (\bibinfo {year} {2007})}\BibitemShut {NoStop}%
\bibitem [{\citenamefont {Shah}(1996)}]{ShahBook96}%
  \BibitemOpen
  \bibfield  {author} {\bibinfo {author} {\bibfnamefont {J.}~\bibnamefont
  {Shah}},\ }\enquote {\bibinfo {title} {Ultrafast spectroscopy of
  semiconductors and semiconductor nanostructures},}\ \ (\bibinfo  {publisher}
  {Springer},\ \bibinfo {address} {Berlin},\ \bibinfo {year} {1996})\
  Chap.~\bibinfo {chapter} {2}\BibitemShut {NoStop}%
\bibitem [{\citenamefont {Gladilin}\ \emph {et~al.}(2004)\citenamefont
  {Gladilin}, \citenamefont {Klimin}, \citenamefont {Fomin},\ and\
  \citenamefont {Devreese}}]{GladilinPRB04}%
  \BibitemOpen
  \bibfield  {author} {\bibinfo {author} {\bibfnamefont {V.~N.}\ \bibnamefont
  {Gladilin}}, \bibinfo {author} {\bibfnamefont {S.~N.}\ \bibnamefont
  {Klimin}}, \bibinfo {author} {\bibfnamefont {V.~M.}\ \bibnamefont {Fomin}}, \
  and\ \bibinfo {author} {\bibfnamefont {J.~T.}\ \bibnamefont {Devreese}},\
  }\href {\doibase 10.1103/PhysRevB.69.155325} {\bibfield  {journal} {\bibinfo
  {journal} {Phys. Rev. B}\ }\textbf {\bibinfo {volume} {69}},\ \bibinfo
  {pages} {155325} (\bibinfo {year} {2004})}\BibitemShut {NoStop}%
\bibitem [{\citenamefont {Mittleman}\ \emph {et~al.}(1994)\citenamefont
  {Mittleman}, \citenamefont {Schoenlein}, \citenamefont {Shiang},
  \citenamefont {Colvin}, \citenamefont {Alivisatos},\ and\ \citenamefont
  {Shank}}]{MittlemanPRB94}%
  \BibitemOpen
  \bibfield  {author} {\bibinfo {author} {\bibfnamefont {D.~M.}\ \bibnamefont
  {Mittleman}}, \bibinfo {author} {\bibfnamefont {R.~W.}\ \bibnamefont
  {Schoenlein}}, \bibinfo {author} {\bibfnamefont {J.~J.}\ \bibnamefont
  {Shiang}}, \bibinfo {author} {\bibfnamefont {V.~L.}\ \bibnamefont {Colvin}},
  \bibinfo {author} {\bibfnamefont {A.~P.}\ \bibnamefont {Alivisatos}}, \ and\
  \bibinfo {author} {\bibfnamefont {C.~V.}\ \bibnamefont {Shank}},\ }\href
  {\doibase 10.1103/PhysRevB.49.14435} {\bibfield  {journal} {\bibinfo
  {journal} {Phys. Rev. B}\ }\textbf {\bibinfo {volume} {49}},\ \bibinfo
  {pages} {14435} (\bibinfo {year} {1994})}\BibitemShut {NoStop}%
\bibitem [{\citenamefont {Dworak}\ \emph {et~al.}(2011)\citenamefont {Dworak},
  \citenamefont {Matylitsky}, \citenamefont {Braun},\ and\ \citenamefont
  {Wachtveitl}}]{DworakPRL11}%
  \BibitemOpen
  \bibfield  {author} {\bibinfo {author} {\bibfnamefont {L.}~\bibnamefont
  {Dworak}}, \bibinfo {author} {\bibfnamefont {V.~V.}\ \bibnamefont
  {Matylitsky}}, \bibinfo {author} {\bibfnamefont {M.}~\bibnamefont {Braun}}, \
  and\ \bibinfo {author} {\bibfnamefont {J.}~\bibnamefont {Wachtveitl}},\
  }\href {\doibase 10.1103/PhysRevLett.107.247401} {\bibfield  {journal}
  {\bibinfo  {journal} {Phys. Rev. Lett.}\ }\textbf {\bibinfo {volume} {107}},\
  \bibinfo {pages} {247401} (\bibinfo {year} {2011})}\BibitemShut {NoStop}%
\bibitem [{\citenamefont {Krauss}\ and\ \citenamefont
  {Wise}(1997)}]{KraussPRL97}%
  \BibitemOpen
  \bibfield  {author} {\bibinfo {author} {\bibfnamefont {T.~D.}\ \bibnamefont
  {Krauss}}\ and\ \bibinfo {author} {\bibfnamefont {F.~W.}\ \bibnamefont
  {Wise}},\ }\href {\doibase 10.1103/PhysRevLett.79.5102} {\bibfield  {journal}
  {\bibinfo  {journal} {Phys. Rev. Lett.}\ }\textbf {\bibinfo {volume} {79}},\
  \bibinfo {pages} {5102} (\bibinfo {year} {1997})}\BibitemShut {NoStop}%
\bibitem [{\citenamefont {Adachi}(2004)}]{AdachiBook04}%
  \BibitemOpen
  \bibinfo {editor} {\bibfnamefont {S.}~\bibnamefont {Adachi}},\ ed.,\
  \href@noop {} {\emph {\bibinfo {title} {Handbook on Physical Properties of
  Semiconductors}}},\ Vol.~\bibinfo {volume} {3}\ (\bibinfo  {publisher}
  {Kluwer Academic},\ \bibinfo {year} {2004})\BibitemShut {NoStop}%
\bibitem [{\citenamefont {Crooker}\ \emph {et~al.}(2003)\citenamefont
  {Crooker}, \citenamefont {Barrick}, \citenamefont {Hollingsworth},\ and\
  \citenamefont {Klimov}}]{CrookerAPL03}%
  \BibitemOpen
  \bibfield  {author} {\bibinfo {author} {\bibfnamefont {S.~A.}\ \bibnamefont
  {Crooker}}, \bibinfo {author} {\bibfnamefont {T.}~\bibnamefont {Barrick}},
  \bibinfo {author} {\bibfnamefont {J.~A.}\ \bibnamefont {Hollingsworth}}, \
  and\ \bibinfo {author} {\bibfnamefont {V.~I.}\ \bibnamefont {Klimov}},\
  }\href {\doibase 10.1063/1.1570923} {\bibfield  {journal} {\bibinfo
  {journal} {Appl. Phys. Lett.}\ }\textbf {\bibinfo {volume} {82}},\ \bibinfo
  {pages} {2793} (\bibinfo {year} {2003})}\BibitemShut {NoStop}%
\bibitem [{\citenamefont {Andreani}(1995)}]{AndreaniBook94}%
  \BibitemOpen
  \bibfield  {author} {\bibinfo {author} {\bibfnamefont {L.~C.}\ \bibnamefont
  {Andreani}},\ }\enquote {\bibinfo {title} {Confined electrons and photons:
  New physics and applications},}\ \ (\bibinfo  {publisher} {Plenum Press},\
  \bibinfo {address} {New York},\ \bibinfo {year} {1995})\ pp.\ \bibinfo
  {pages} {57--112}\BibitemShut {NoStop}%
\bibitem [{\citenamefont {Langbein}\ \emph {et~al.}(2002)\citenamefont
  {Langbein}, \citenamefont {Mann}, \citenamefont {Woggon}, \citenamefont
  {Klude},\ and\ \citenamefont {Hommel}}]{LangbeinPSSA02a}%
  \BibitemOpen
  \bibfield  {author} {\bibinfo {author} {\bibfnamefont {W.}~\bibnamefont
  {Langbein}}, \bibinfo {author} {\bibfnamefont {C.}~\bibnamefont {Mann}},
  \bibinfo {author} {\bibfnamefont {U.}~\bibnamefont {Woggon}}, \bibinfo
  {author} {\bibfnamefont {M.}~\bibnamefont {Klude}}, \ and\ \bibinfo {author}
  {\bibfnamefont {D.}~\bibnamefont {Hommel}},\ }\href@noop {} {\bibfield
  {journal} {\bibinfo  {journal} {phys. stat. sol. (a)}\ }\textbf {\bibinfo
  {volume} {190}},\ \bibinfo {pages} {861} (\bibinfo {year}
  {2002})}\BibitemShut {NoStop}%
\bibitem [{\citenamefont {Wagner}\ \emph {et~al.}(1998)\citenamefont {Wagner},
  \citenamefont {Sch\"atz}, \citenamefont {Maier}, \citenamefont {Langbein},\
  and\ \citenamefont {Hvam}}]{WagnerPRB98}%
  \BibitemOpen
  \bibfield  {author} {\bibinfo {author} {\bibfnamefont {H.~P.}\ \bibnamefont
  {Wagner}}, \bibinfo {author} {\bibfnamefont {A.}~\bibnamefont {Sch\"atz}},
  \bibinfo {author} {\bibfnamefont {R.}~\bibnamefont {Maier}}, \bibinfo
  {author} {\bibfnamefont {W.}~\bibnamefont {Langbein}}, \ and\ \bibinfo
  {author} {\bibfnamefont {J.~M.}\ \bibnamefont {Hvam}},\ }\href {\doibase
  10.1103/PhysRevB.57.1791} {\bibfield  {journal} {\bibinfo  {journal} {Phys.
  Rev. B}\ }\textbf {\bibinfo {volume} {57}},\ \bibinfo {pages} {1791}
  (\bibinfo {year} {1998})}\BibitemShut {NoStop}%
\bibitem [{\citenamefont {Dolan}\ \emph {et~al.}(2010)\citenamefont {Dolan},
  \citenamefont {Hughes}, \citenamefont {Grazioso}, \citenamefont {Patton},\
  and\ \citenamefont {Smith}}]{DolanOL10}%
  \BibitemOpen
  \bibfield  {author} {\bibinfo {author} {\bibfnamefont {P.~R.}\ \bibnamefont
  {Dolan}}, \bibinfo {author} {\bibfnamefont {G.~M.}\ \bibnamefont {Hughes}},
  \bibinfo {author} {\bibfnamefont {F.}~\bibnamefont {Grazioso}}, \bibinfo
  {author} {\bibfnamefont {B.~R.}\ \bibnamefont {Patton}}, \ and\ \bibinfo
  {author} {\bibfnamefont {J.~M.}\ \bibnamefont {Smith}},\ }\href {\doibase
  10.1364/OL.35.003556} {\bibfield  {journal} {\bibinfo  {journal} {Opt.
  Lett.}\ }\textbf {\bibinfo {volume} {35}},\ \bibinfo {pages} {3556} (\bibinfo
  {year} {2010})}\BibitemShut {NoStop}%
\bibitem [{\citenamefont {Schnabel}\ \emph {et~al.}(1992)\citenamefont
  {Schnabel}, \citenamefont {Zimmermann}, \citenamefont {Bimberg},
  \citenamefont {Nickel}, \citenamefont {L{\"o}sch},\ and\ \citenamefont
  {Schlapp}}]{SchnabelPRB92}%
  \BibitemOpen
  \bibfield  {author} {\bibinfo {author} {\bibfnamefont {R.~F.}\ \bibnamefont
  {Schnabel}}, \bibinfo {author} {\bibfnamefont {R.}~\bibnamefont
  {Zimmermann}}, \bibinfo {author} {\bibfnamefont {D.}~\bibnamefont {Bimberg}},
  \bibinfo {author} {\bibfnamefont {H.}~\bibnamefont {Nickel}}, \bibinfo
  {author} {\bibfnamefont {R.}~\bibnamefont {L{\"o}sch}}, \ and\ \bibinfo
  {author} {\bibfnamefont {W.}~\bibnamefont {Schlapp}},\ }\href@noop {}
  {\bibfield  {journal} {\bibinfo  {journal} {Phys. Rev. B}\ }\textbf {\bibinfo
  {volume} {46}},\ \bibinfo {pages} {9873} (\bibinfo {year}
  {1992})}\BibitemShut {NoStop}%
\bibitem [{\citenamefont {Pasquarello}\ and\ \citenamefont
  {Andreani}(1991)}]{PasquarelloPRB91}%
  \BibitemOpen
  \bibfield  {author} {\bibinfo {author} {\bibfnamefont {A.}~\bibnamefont
  {Pasquarello}}\ and\ \bibinfo {author} {\bibfnamefont {L.~C.}\ \bibnamefont
  {Andreani}},\ }\href@noop {} {\bibfield  {journal} {\bibinfo  {journal}
  {Phys. Rev. B}\ }\textbf {\bibinfo {volume} {44}},\ \bibinfo {pages} {3162}
  (\bibinfo {year} {1991})}\BibitemShut {NoStop}%
\bibitem [{\citenamefont {Tessier}\ \emph {et~al.}(2013)\citenamefont
  {Tessier}, \citenamefont {Biadala}, \citenamefont {Bouet}, \citenamefont
  {Ithurria}, \citenamefont {Abecassis},\ and\ \citenamefont
  {Dubertret}}]{TessierACSN13}%
  \BibitemOpen
  \bibfield  {author} {\bibinfo {author} {\bibfnamefont {M.~D.}\ \bibnamefont
  {Tessier}}, \bibinfo {author} {\bibfnamefont {L.}~\bibnamefont {Biadala}},
  \bibinfo {author} {\bibfnamefont {C.}~\bibnamefont {Bouet}}, \bibinfo
  {author} {\bibfnamefont {S.}~\bibnamefont {Ithurria}}, \bibinfo {author}
  {\bibfnamefont {B.}~\bibnamefont {Abecassis}}, \ and\ \bibinfo {author}
  {\bibfnamefont {B.}~\bibnamefont {Dubertret}},\ }\href {\doibase
  10.1021/nn400833d} {\bibfield  {journal} {\bibinfo  {journal} {ACS Nano}\
  }\textbf {\bibinfo {volume} {7}},\ \bibinfo {pages} {3332} (\bibinfo {year}
  {2013})}\BibitemShut {NoStop}%
\end{thebibliography}
%

\end{document}